# *Ab initio* studies of phonon softening and high pressure phase transitions of α-quartz $SiO_2$

**N. Choudhury and S. L. Chaplot**
Solid State Physics Division, Bhabha Atomic Research Centre
Trombay, Mumbai 400 085, India

***Abstract***

Density functional perturbation theory calculations of α-quartz using extended norm conserving pseudopotentials have been used to study the elastic properties and phonon dispersion relations along various high symmetry directions as a function of bulk, uniaxial and non-hydrostatic pressure. The computed equation of state, elastic constants and phonon frequencies are found to be in good agreement with available experimental data. A zone boundary (1/3, 1/3, 0) K-point phonon mode becomes soft for pressures above P=32 GPa. Around the same pressure, studies of the Born stability criteria reveal that the structure is mechanically unstable. The phonon and elastic softening are related to the high pressure phase transitions and amorphization of quartz and these studies suggest that the mean transition pressure is lowered under non-hydrostatic conditions. Application of uniaxial pressure, results in a post-quartz crystalline monoclinic *C2* structural transition in the vicinity of the K-point instability. This structure, intermediate between quartz and stishovite has two-thirds of the silicon atoms in octahedral coordination while the remaining silicon atoms remain tetrahedrally coordinated. This novel monoclinic *C2* polymorph of silica, which is found to be metastable under ambient conditions, is possibly one of the several competing dense forms of silica containing octahedrally coordinated silicon. The possible role of high pressure ferroelastic phases in causing pressure induced amorphization in silica are discussed.

## I. Introduction

The silica polymorph quartz exhibits several interesting properties[1-17] including pressure induced amorphization[1-2, 8], high pressure and temperature phase transitions[3-4, 7, 11-19], anomalous elastic properties[3,6], negative Poisson ratios[6(b-c)], soft mode behavior[9-10, 17], *etc.* Silica in its various crystalline and amorphous forms finds several industrial applications including being a raw material for glasses, ceramics, production of silicon, *etc.* Quartz oscillators and optical waveguides are used extensively in long distance telecommunications and industry. Despite its simple chemical composition, silica is known to assume various structures which have a wide range of densities (2.3-4.3 gm/cm$^3$) and bulk modulii and has aroused considerable theoretical[7-26] and experimental[1,3-6(a-b),27-30] interest.

α-quartz is the most stable silica polymorph at ambient conditions (up to 3 GPa), and persists as a metastable state at higher pressures. Quartz amorphizes at pressures of around 18-35 GPa[1-3]. On release of pressure, quartz remains amorphous[1], but is anisotropic with memory of the quartz crystallographic orientation[1(c),6(a)]. The atomic disorder with pressure was therefore believed to derive from small perturbations of the quartz structure. Although pressure induced amorphization occurs in a variety of solids like α-quartz[1-2], coesite[22], ice[31], *etc.* its origins are not clearly understood. Several post-quartz crystalline phases have also been reported both from theory[7, 12-14, 16] and experiments[1(b), 4]. First principles molecular dynamics simulations predicted a new structure for silica which results from annealing quartz at pressures near a major phonon instability[7]. The diffraction pattern of this phase compared favorably with that of an unidentified intermediate crystalline phase[1(b)] found in silica during the amorphization.

Molecular dynamics (MD) simulations using interatomic potentials fitted to first principles total energy surfaces[24-25] reveal pressure induced amorphization[8] at P~ 22 GPa, in agreement with experiments. Around the amorphization pressure, α-quartz displays soft phonon modes[9-10] with a K-point phonon mode becoming unstable above P=21 GPa. First principles high pressure calculations, however indicated that the K-point phonon instability occurs at a higher pressure[17]. Although the interatomic potentials[24-26] reproduce the observed pressure induced amorphization transition[9-10] pressure correctly, there are important differences in the computed elastic constants and Born stability criteria[15], Raman and infrared data as compared to experimental data[3,29-30] and available first principles results[15,20].

We report density functional theory[32-38](DFT) and density functional perturbation theory[32-38] (DFPT) calculations of the equation of state, elastic constants, Born stability criteria, phonon dispersion relations and their variations with pressure of α-quartz. Our goals are to (i) understand the K-point phonon instability and its variations under bulk and uniaxial compression as well as under non-hydrostatic conditions (ii) study the high pressure phase transitions of quartz, and (iii) understand the mechanical stability of quartz at high pressure as studied from the elastic constants and Born stability criteria. There are various controversies in the literature[3,16] on the appropriate Born stability criteria to be applied to study the mechanical stability of quartz at high pressure, *etc.* We address these issues, and have undertaken systematic high pressure studies of a variety of properties, including under non-hydrostatic conditions. A new high pressure crystalline post-quartz monoclinic phase is obtained and its structural properties studied.

## II. Techniques

Quartz has a trigonal structure[28(e)] (space group *P3$_2$21*) with 9 atoms/unit cell. Density functional perturbation theory (DFPT) linear response calculations with a plane-wave basis set as implemented in the code ABINIT[39] were undertaken. We used the local density approximation (LDA) involving a rational polynomial parametrization of the exchange-correlational functional[40] based on Ceperley and Alder[41] electron-gas data and employ well tested extended norm conserving separable pseudopotentials[42] used earlier for zero pressure calculations of α-quartz[20] and stishovite[21]. The computed phonon frequencies and LO-TO splittings[20(a), 21], dispersion relations[20(b), 21], density of states[20(c)], specific heat[20(c)] and mean square atomic displacements[20(c)] using these pseudopotentials were found to be in good agreement with experiments. A large planewave energy cutoff of 120 Ry and smearing of 1 Ry were used to obtain precise values of the stress tensor in the high pressure studies. The Brillouin zone (BZ) integrations were performed using a 4×4×4 special **k**-point mesh. Selected computations of the long wavelength and zone boundary K-point phonons with a 6×6×6 special k-point mesh showed that the 4x4x4 mesh was sufficient and the total energies, structural variables and phonon frequencies were well converged at the highest pressures. At each pressure, the structures were completely relaxed prior to the calculations.

To obtain the phonon frequencies at a general wave vector, we first computed the dynamical matrices on the (4,4,4) wavevector grid involving 10 wave vectors in the reciprocal space and computed the interatomic force constants. The long-range dipole-dipole interactions were taken into account separately using the calculated anisotropic Born effective charge tensor and dielectric tensor. The elastic constants as a function of pressure were evaluated both from DFPT linear response[38] using a 6x6x6 special **k**-point mesh with an energy cutoff of 120 Ry as well as directly from the stress-strain relations[43]. To compute the elastic constants under pressure, we applied the appropriate strains[44] that yield the various elastic constants and computed the total energies and stresses, after careful relaxation of the strained structures. Linear response DFPT studies with strain perturbations using the reduced coordinate metric tensor method of Hamann *et al.*[38] were also used to study the elastic constants at high pressure. The linear response results give the proper elastic constant tensors[38] and appropriate pressure corrections[38, 45-47] were applied to obtain the stiffness constants which satisfy the Born stability criteria.

## III. Results

### (A) Equation of state and long wavelength phonon frequencies

The computed crystal structure, equation of state and long wavelength phonon frequencies (Figs. 1-3) are found to be in good agreement with reported high pressure experimental data[28] and available first principles LDA calculations[20(a), 48-50]. First principles LDA[48-50] and Generalized Gradient Approximation (GGA)[49-50] studies of the equation-of-state of α-Quartz have been reported by various workers[48-50] and we have compared our calculated results with the theoretical[49] and reported experimental[28(a-d)] studies in Fig. 2. Quartz has a framework structure (Fig. 1) formed by corner shared silicate tetrahedral units. Application of pressure distorts and causes tilts in the tetrahedral units and quartz reconstructively transforms successively to the coesite and octahedrally coordinated stishovite phases at high pressure. Due to large kinetic barriers, quartz however persists as a metastable state up to very high pressures at room temperature. The 4 to 6 coordinated silicon transition is known to cause seismic discontinuities in the Earth's mantle which marks the transition from the upper to lower mantle, and understanding the mechanism of this transition particularly for a model system like silica is particularly useful. Hamann's studies[50] on the quartz to stishovite phase transition indicates that only the GGA yields correct relative energies and leads to a reasonably accurate prediction of the transition pressure. However the structural parameters within the GGA are less accurately predicted[50]; these have motivated the development of more accurate GGA[51]. The LDA calculations[48-50] of α-quartz predict the equilibrium volume with an accuracy of 1%, whereas the GGA[49,50] lead to an overestimate of 6.6%. The LDA also lead to a consistently more accurate description of all atomic coordinates[48,50]; we have therefore adopted the LDA for our high pressure studies. Even within the LDA, the reported bulk modulii show a large scatter (35 GPa[49], 45 GPa[50], both having B'=4.9; B=38.1 GPa[48]); the corresponding equilibrium LDA volumes[48-50] typically scatter between 35.8 $Å^3$ and 38.55 $Å^3$. Experimental synchrotron x-ray diffraction measurements of Haines *et. al*,[4] indicate that quartz is metastable upto much higher pressures than the reported equation-of-state measurements[28] which are typically available in the 0-10 GPa range; in the present study, we report the equation-of-state over the entire pressure range (0-32 GPa) over which α-quartz is dynamically and mechanically stable (Fig. 2).

The long wavelength phonon modes in α-quartz at the zone center Γ point can be classified as

$$G:\ 4A_1 + 5A_2 + 9E$$

The $E$ phonon modes are doubly degenerate polar phonon modes. The computed structures and variations of the Born effective charge tensors, electronic dielectric tensors and phonon frequencies with pressure are given in Tables I-V. All the structures given in Table I are dynamically stable in the entire Brillouin zone and represent stable and metastable high pressure quartz structures under hydrostatic and non-hydrostatic conditions. Both the Born effective charges (Table II) and electronic dielectric tensors (Table III) are required to understand the influence of the macroscopic electric field on long wavelength phonon properties. The Born effective charge tensor gives a measure of the local dipole moment which develops when the nuclei are moved and corresponds to the variation of the polarization with atomic displacements. As the Born effective charge tensor $Z^*_{ab}(k)$ is a mixed second derivative of the total energy, with respect to macroscopic electric field component $E_a$ and atomic displacement component $\tau_{k,b}$, there is no requirement that the tensor be symmetric. The pressure variations of the low and high frequency dielectric tensors are given in Table III. The zero pressure dielectric tensors are overestimated as is typical within the LDA; however, application of the scissors correction[20(a)] has been found to improve the agreement with experimental data for both high and low frequency tensors. Gonze *et al.*[20(a)] discuss the important role of the anisotropy of the Born effective charge tensors in describing the observed zero pressure LO-TO splittings in α-quartz. Our zero pressure results (Table IV) are in complete agreement with their studies[20(a)]. At higher pressures, the anisotropy of the charge tensor (Table III) is found to significantly reduce which indicates important changes in bonding character. The LO-TO splittings of the $A_2$ phonon modes (with macroscopic electric field parallel to the c-axis) especially the 343 and 763 $cm^{-1}$ phonon modes, significantly increase with pressure. The splittings of the doubly degenerate $E$ phonon modes which have their macroscopic electric field perpendicular to the $c$-axis are not largely altered by pressure with the simultaneous hardening of both the LO and TO modes. The 465 $cm^{-1}$ non-polar $A_1$ phonon mode (Table V) hardens significantly with pressure in good agreement with experimental measurements[29].

**(B) High pressure elastic constants and Born stability criteria**

The computed $P=0$ elastic constants obtained from linear response DFPT calculations are in good agreement with reported experimental data (Table VI). We have compared our zero pressure elastic constants with the calculated LDA results of Holm and Ahuja[18]; although the overall agreements with experiments seem similar, the calculated individual elastic constants obtained have some differences (Table VI). The elastic constants of α-quartz , particularly, the soft elastic constants $C_{12}$, $C_{13}$ and $C_{14}$ are very sensitive to the relaxed structural parameters obtained. While we obtain better agreement with experiments for $C_{13}$ and $C_{14}$, Holm and Ahuja[18] obtain better agreement for $C_{11}$ and $C_{12}$. It is interesting to see that even in the experiments, the $C_{33}$ values reported[3,5] (Table VI) are different; a somewhat large scatter is also obtained in the bulk modulii and observed equation of state[28(a-d)] .

There are various definitions of the elastic constants[38,43,45-47]; however the elastic constant obtained from stress-strain relations and from long-wavelength lattice dynamics are different[43,45-47] for non-zero stresses. The elastic constants that define the mechanical stability criteria under pressure are those appearing in stress-strain relations[45-47]. The linear response formulation[32] taking into account strain perturbations on the other hand gives the "proper elastic constant tensor"[38]. While these two definitions become identical at zero pressure, under hydrostatic pressure appropriate pressure corrections have to be applied[38, 45-47] to the linear response results.

The computed high pressure elastic constants of α-quartz evaluated for bulk hydrostatic pressure using the stress-strain relations are compared with reported high pressure Brillouin scattering measurements[4] in Fig. 4. The computed elastic constants obtained from linear response studies with strain perturbations and appropriate pressure corrections[38, 45-47] are found to be in excellent agreement (to well within 0.5 GPa) with the values obtained from stress-strain relations. The elastic constants $C_{11}$ and $C_{33}$ increase sharply with pressure. Although the zero pressure elastic constants are in good agreement with experiments (Table VI), the computed high pressure average elastic constants seem underestimated (Fig. 4). The elastic constants of α-quartz are found to be quite sensitive to the hydrostatic pressure conditions and the variations could perhaps be due to the difficulties in maintaining perfect hydrostatic conditions at high pressure in the experiments.

The mechanical Born stability criteria for trigonal structures results in various constraints for the elastic constants which are given by[3,11(a),15(b)]

$$B_1 = C_{11}-|C_{12}| > 0$$
$$B_2=(C_{11}+C_{12})C_{33}-2(C_{13})^2 > 0 \qquad (1).$$
$$B_3=(C_{11}-C_{12})C_{44}-2(C_{14})^2 > 0$$

The computed variations of $B_1$, $B_2$ and $B_3$ with pressure are shown in Fig. 5. All the three conditions given in Equation (1) must be simultaneously satisfied for the system to be mechanically stable[11(a),15]. $B_1$ and $B_2$ are positive up to 40 GPa (Fig. 5).

The quartz structure becomes mechanically unstable around P=32 GPa as $B_3$ becomes negative and the Born stability criteria gets violated. Our computed results and trends for $B_1$, $B_2$ and $B_3$ (Fig. 5 ) are in good qualitative agreement with the reported experimental Brillouin scattering results of Gregoryanz *et al.*[3] and the theoretical studies of Bingelli *et al.*[15(b)]. While the interatomic potentials yield lower transition pressures[15(b)] of around P=20 GPa, the first principles results reveal that the system is mechanically unstable at pressures of around P=32 GPa.

**(C) . High pressure phonon instabilities and structural phase transitions**

The computed phonon dispersion relations in α-quartz at various bulk pressures, namely, P=0, 24 , 30 and 38 GPa are shown in Fig. 6. Almost all the zone center Raman and infrared modes harden with pressure as observed experimentally (Fig. 3). The zero pressure phonon dispersion reveals a large band gap from 24-31 THz; this gap significantly lowers at high pressures. The lowest zone boundary K-point phonon mode becomes soft for bulk pressures above P=32 GPa and the pressure evolution of the phonon instability can be clearly seen from Figs. 5 and 6. At P=24 GPa, the K-point lowest phonon mode is quite stable; the onset of the instability occurs well above this pressure.

We have also computed the "elastic constant" corresponding to the soft acoustic mode using the relation[3]

$$r v^2 = \frac{1}{4}\{(C_{11} - C_{12} + 2C_{44}) - [(C_{11} - C_{12} - 2C_{44})^2 + 16C_{14}^2]^{\frac{1}{2}}\} \qquad (2).$$

For small $B_3$, $r v^2 = \frac{B_3 C_{44}}{2[C_{14}^2 + C_{44}^2]}$, which is found to be proportional to $B_3$. The computed $r v^2$ (Fig. 5) goes through a maximum near 6.5 GPa and then decreases linearly, vanishing at high pressures. Similar trends have been noticed from Brillouin scattering measurements[3] and have been associated with proper ferroelastic behavior in the high pressure regime[3]. The phase transition pressure obtained from the vanishing "elastic constant" $r v^2$ (Fig. 5(d)) and K-point phonon frequency (Fig. 5(d)) are very close (P~32 GPa). The transition pressure predicted from first principles studies (P=32 GPa) is higher than that obtained from interatomic potentials (P=21 GPa[9] using Tsuneyuki *et al.* potentials[24], P=27 GPa[10] using other potentials) and in good agreement with the LDA calculations of Baroni and Giannozzi[17] giving the K-point instability at P~32 GPa. The mechanical instability of the quartz structure obtained from the Born stability criteria $B_3$ occurs in the vicinity of the K-point instability, both using interatomic potentials[15b] and from first principles studies.

In Fig. 6, we also display the phonon dispersion for various non-hydrostatic situations. While the plot in Fig. 6(e) with non-hydrostatic pressure involving high pressure along the *c*-axis as well as the *ab*-plane has a dispersion similar to under hydrostatic situations, there is a drastic change in dispersion for compression only along the *c*-axis (Fig. 6(f)). On gradual application of uniaxial pressure, phonon softening is found for the stress tensor $\sigma_{xx}=\sigma_{yy}=0$, $\sigma_{zz}=-15.8$ GPa, $\sigma_{zx}=\sigma_{xy}=\sigma_{yz}=0$. The phonon softening is related to the high pressure phase transitions and amorphization of quartz[9-10] and these results suggest that the mean transition pressure is significantly lowered under non-hydrostatic conditions. This possibly explains the wide range of amorphization pressure (18-35 GPa) obtained in the experiments[1-3]. The nature of phonon dispersion (Fig. 6(f)) along the Γ-A direction is quite different for uniaxial compression along the *c*-axis.

Unconstrained structural relaxation of the 9-atom unit cell of α-quartz around this stress tensor $\sigma_{xx}=\sigma_{yy}=0$, $\sigma_{zz}=-15.8$ GPa, $\sigma_{zx}=\sigma_{xy}=\sigma_{yz}=0$ yields a post-quartz transition to a crystalline monoclinic *C2* structure. Gradual symmetry preserving structural relaxation yields the corresponding monoclinic *C2* structure for bulk pressures. The *C2* structure has a significantly lower volume which results in a lower enthalpy (which gives the zero temperature free energy, neglecting zero-point vibrational contributions) than the α-quartz structure for bulk pressures of above P=9 GPa (Fig. 7). Two thirds of the silicon atoms are in octahedral coordination in the *C2* structure (Table VII), while the remaining one-thirds silicon are tetrahedrally coordinated both under uniaxial and bulk pressure conditions. The *C2* structure has edge-shared octahedral units with corner shared silicate tetrahedra which form a framework structure (Fig.1).

The computed x-ray diffraction patterns and *d*-spacings of the post-quartz monoclinic *C2* structure at P=38 GPa have been studied (Table VIII). The largest intensities in the *C2* structure occurs for *d*-spacings (Å) of 4.7944 (19.5), 3.6912 (14.5), 3.0573(26.7), 2.7505 (50), 2.6045 (64.8), 2.4478 (31.4), 1.8456 (39.67), 1.7819 (33.35) and 1.6642 (30.91), where the numerical values in parenthesis give the relative intensities. The computed x-ray intensities of the high pressure *C2* structure are compared with experimental synchrotron x-ray diffraction data (Fig. 8) for the observed high pressure post-quartz crystalline phases[1(b),4]. Haines *et al.*[4], obtained a post-quartz crystalline phase by compressing quartz to 45 GPa at room temperature in a close to hydrostatic, helium pressure medium. Kingma *et al.*[1(b)] have reported the energy-dispersive diffraction spectra of a post-quartz crystalline phase obtained by quasi-hydrostatic compression of polycrystalline α-quartz. The computed diffraction patterns (Fig. 8) and *d*-spacings (Table VIII) of the *C2* structure are overall in fair agreement with the synchrotron x-ray data of Haines *et al.*,[4] which have been indexed based on a low symmetry $P2_1/c$ monoclinic structure[4]. While we obtain diffraction peaks and *d*-spacings around the positions indicated by Kingma *et al.*[1(b)], the *C2* structure has more peaks (Table VIII, Fig. 8) and lower symmetry than their reported post-quartz crystalline phase[1(b)].

Although the monoclinic *C2* structure has a lower free energy for pressures above P=9 GPa (Fig. 7), quartz can persist metastably up to the point of onset of elastic and dynamical instability, which occurs around bulk pressures of P=32 GPa (Fig. 5). This novel *C2* monoclinic polymorph of silica, which is found to be metastable under ambient conditions, is possibly one of the several competing[23] dense forms of silica containing octahedrally coordinated silicon. It is interesting to note that the zero pressure metastable *C2* quenched structure obtained from our studies has two-thirds of the silicon atoms which are 5 coordinated to within 1.8 Å, with a sixth oxygen at 2.04 Å. The longest Si-O bond is quite compressible, and with increasing bulk pressure, the octahedral distortions in the *C2* structure are lowered.

**IV. Discussion**

First principles calculations have revealed that the energy landscape in silica at high pressure is quite complex and there are several possible octahedrally coordinated competing dense high pressure silica structures[23]. The crystalline high pressure structures obtained seems to depend on the kinetics and the pathway adopted and several post quartz crystalline structures have been reported both from theory[7,12,14,16] and experiments[1,4]. Haines *et al.*[4] have indexed the diffraction patterns for their observed high pressure crystalline post quartz phase based on a monoclinic cell with space group $P2_1/c$, with a model structure built up of 3x3x2 zigzag chains of $SiO_6$ octahedra. There are key differences in the observed high pressure post-quartz crystalline structures of Kingma *et al.*[1] and Haines *et al.*[4], which probably stem from the different hydrostatic conditions and pathways adopted in the experiments.

First principles molecular dynamics simulations[7] predicted a new trigonal structure for silica (space group $P3_2$) with two thirds of silicon atoms in octahedral coordination and one third which are 5-coordinated. The diffraction pattern of this phase compared well with the crystalline phase reported by Kingma *et al.*[1]. Due to the different approaches adopted, the post-quartz crystalline structures obtained by Wentzcovitch *et al.*[7] are different from those obtained in the present study. While our results involve structural optimization of the 9 atom quartz unit cell at high pressure and zero-temperature, Wentzcovitch *et al.*[7], optimized the geometry of their 27 atom supercell while compressing it to 33 GPa and their resulting structure was annealed at temperatures fluctuating between 300–600 K for 0.8 ps followed by a rapid quench to 0 K.

Theoretical calculations using interatomic potentials to understand the high pressure post quartz crystalline phases have also been reported[12-14,16]. Molecular-dynamics calculations[12] at T=300 K reveal a crystalline-to-crystalline transition from α-quartz to a phase with five-coordinated silicon structure at high pressure under non-hydrostatic conditions having the same space group as α-quartz. Various other low symmetry monoclinic structures[14,16] have also been obtained using the Tsuneyuki *et al.*[24] and Van Beest *et al.*[25] interatomic potentials. Our studies reveal that while the simulations using interatomic potentials fitted to first principles energy surface give results in qualitative agreement with first principles, there are important differences. A possible reason for this discrepancy is the inability of isotropic pair potentials to accurately describe the zero pressure LO-TO splittings[20] and in consequence the complete phonon spectra of quartz[20]. There are reported controversies[3,16] on the appropriate Born stability criteria adopted at high pressure and the reported mechanical instabilities at high pressure. Our studies however, are in qualitative agreement with the reported Brillouin scattering measurements of Gregoryanz *et al*[3(a)] that suggest high pressure ferroelastic transitions in α-quartz driven by the violations of the Born stability criteria for $B_3$.

The new *C2* structure we report as well as the various high pressure post-quartz structures reported by others[7,16] are interestingly racemic with chiral properties. The α-quartz structure features helices of corner-linked $SiO_4$ tetrahedra that can adopt either left- or right-handed configurations; the chiral behavior of quartz has aroused much interest[53,54]. Theoretical studies using interatomic potentials[16] indicate that while the high pressure monoclinic quartz II structure is ferroelastic in principle, the transition itself is coelastic, as the shape of the newly formed crystal is determined by the handedness of α-quartz. *Ab initio* studies of the energy barriers for switching the handedness in the high pressure monoclinic quartz structure are desirable, due to the differences between the interatomic potentials and first principles results.

Toledano *et al.*[55] have explained the mechanism of amorphization under pressure of α-quartz and ice, and the microstructural properties of some ferroelectric relaxors based on a ferroelastic glass state which may form in a crystalline material under the following conditions[55] (i) the ferroelastic structure can exist in crystallographic configurations corresponding to different spontaneous strain components, and, (ii) an internal stress field is created involving the stress components conjugated to the preceding strains which induces sufficiently large mismatches between the differently sheared domains, leading to a splitting and disintegration of the mesoscopic-size domains into nanodomains, destroying the long-range order in the crystal.

Ferroelastic behavior in the high pressure regime in α-quartz is indicated and is of interest as spontaneous strains can lower the symmetry. All these materials including silica[23], ice[31] and some relaxors[56] seem to have a complex energy landscape with a soft energy surface and various competing energetically equivalent structures. It is interesting to contemplate whether competing high pressure ferroelastic phases can lead to a ferroelastic glass state in silica which in turn leads to its amorphization. A microscopic understanding of the complex energy landscape in silica and the possible role of ferroelastic phases may perhaps provide more insights into the fundamental mechanisms of the observed high pressure amorphization.

## V. Conclusions

High pressure first principles density functional theory calculations help understand the role of phonon and elastic instabilities in the high pressure structural phase transitions and amorphization of α-quartz. Non-hydrostatic stresses are found to significantly influence the instabilities and explain the wide range of transition pressures observed in the experiments. A new high pressure post quartz dense silica phase is obtained from structural relaxation in the vicinity of the K-point instability in which two-thirds of the silicon atoms are in octahedral coordination. This structure with effective silicon coordination intermediate between quartz and stishovite is metastable at ambient pressure. Possible links between the ferroelastic nature of high pressure silica transformations and pressure induced amorphization are outlined.

*Acknowledgements*
NC acknowledges useful discussions and partial support from R.E Cohen of the Carnegie Institution of Washington (National Science Foundation grant EAR-0310139 to R.E.C. ). NC also thanks X.Gonze and F. Detraux for useful discussions These studies have been carried out using the supercomputing resources of the Bhabha Atomic Research Centre (BARC), India. We gratefully acknowledge computing support from Computer Division, BARC and we thank K. Bhatt and R.S. Sharma, RRSD for technical support..

**Table I:** Stable and metastable quartz structures as a function of pressure obtained from symmetry preserving structural relaxations with space group $P3_221$. The *S*i atoms are in *3b* Wyckoff positions with fractional coordinated *(u, 0, 1/6 ), (0, u, 5/6 ), ( - u, - u, 1/2 )* and the *O* atoms in *6c* positions with coordinates *(x, y, z) ( - y, x - y, z + 2/3 ) ( - x + y, - x, z + 1/3 ) (y, x, - z) (x - y, - y, - z + 1/3 ) ( - x, - x + y, - z + 2/3 )*. *P1*, *P2* and *P3* represent non-hydrostatic pressures with stress tensor $\sigma_{xx}=\sigma_{yy}$=-19 GPa, $\sigma_{xy}= \sigma_{yz}= \sigma_{zx}$=0 and $\sigma_{zz}$= -24.8, -29.6 and -38 GPa, respectively. *Pc1* has the stress tensor $\sigma_{xx}=\sigma_{yy}$=-1.1 GPa, $\sigma_{xy}= \sigma_{yz}= \sigma_{zx}$=0 and $\sigma_{zz}$= -6.9 GPa. All these structures are dynamically stable in the entire Brillouin zone.

| | P (GPa) | *a* (Å) | *c* (Å) | *u* | *x* | *y* | *z* |
|---|---|---|---|---|---|---|---|
| Bulk Pressure | | | | | | | |
| | 0.24 | 4.8137 | 5.3150 | 0.4613 | 0.4097 | 0.2814 | 0.2748 |
| | 7.1 | 4.5795 | 5.1908 | 0.4451 | 0.3964 | 0.3048 | 0.2589 |
| | 19.1 | 4.3444 | 5.0881 | 0.4311 | 0.3773 | 0.3208 | 0.2513 |
| | 23.7 | 4.2773 | 5.0584 | 0.4271 | 0.3708 | 0.3247 | 0.2501 |
| | 29.6 | 4.2057 | 5.0229 | 0.4227 | 0.3633 | 0.3286 | 0.2491 |
| Non-hydrostatic Pressure | | | | | | | |
| | P1 | 4.3825 | 4.9724 | 0.4286 | 0.3780 | 0.3203 | 0.2494 |
| | P2 | 4.4135 | 4.8887 | 0.4271 | 0.3788 | 0.3193 | 0.2484 |
| | P3 | 4.4813 | 4.7386 | 0.4258 | 0.3809 | 0.3161 | 0.2468 |
| | Pc1 | 4.8316 | 5.0438 | 0.4546 | 0.4070 | 0.2879 | 0.2646 |

**Table II**: Variations of the dynamic Born effective charge tensors of the Si and O atoms with pressure. Only the tensors of the inequivalent atoms are listed. The charge tensors of other atoms can be derived from symmetry. *x* and *z* are respectively, along the *a* and *c*-axis.

| | P (GPa) | $Z^*_{xx}$ | $Z^*_{yy}$ | $Z^*_{zz}$ | $Z^*_{xy}$ | $Z^*_{xz}$ | $Z^*_{yx}$ | $Z^*_{yz}$ | $Z^*_{zx}$ | $Z^*_{zy}$ |
|---|---|---|---|---|---|---|---|---|---|---|
| Si | 0.24 | 3.01 | 3.63 | 3.45 | 0 | 0 | 0 | .28 | 0 | -.32 |
| O | 0.24 | -1.32 | -2.00 | -1.72 | .42 | .22 | .48 | -.71 | .29 | -.66 |
| Si | 19.1 | 3.17 | 3.53 | 3.53 | 0 | 0 | 0 | .26 | 0 | -.32 |
| O | 19.1 | -1.39 | -1.96 | -1.76 | .31 | .043 | .36 | -.56 | .11 | -.48 |
| Si | 29.6 | 3.25 | 3.56 | 3.58 | 0 | 0 | 0 | .21 | 0 | -.31 |
| O | 29.6 | -1.43 | -1.97 | -1.79 | .28 | .00 | .33 | -.51 | .05 | -.42 |
| Si | 38.4 | 3.31 | 3.61 | 3.63 | 0 | 0 | 0 | .16 | 0 | -.30 |
| O | 38.4 | -1.57 | -1.99 | -1.81 | .27 | -0.02 | .31 | -.48 | .01 | -.38 |

**Table III**: The computed electronic ($\varepsilon^{\infty}$) and zero frequency ($\varepsilon^0$) dielectric tensors as function of bulk pressure. *z* and *x* are respectively, along and perpendicular to the *c*-axis.

| | P=.24 GPa | P=7 GPa | P=19 GPa | P=24 GPa | P=30 GPa |
|---|---|---|---|---|---|
| $\varepsilon^{\infty}_{xx}$ | 2.53 | 2.71 | 2.90 | 2.97 | 3.04 |
| $\varepsilon^{\infty}_{zz}$ | 2.56 | 2.75 | 2.95 | 3.01 | 3.08 |
| $\varepsilon^{0}_{xx}$ | 4.766 | 5.19 | 5.87 | 6.16 | 6.55 |
| $\varepsilon^{0}_{zz}$ | 4.967 | 5.37 | 6.03 | 6.31 | 6.70 |

**Table IV**: The computed long wavelength longitudinal optic (LO) and transverse optic (TO) phonon frequencies ($cm^{-1}$) and LO-TO splittings ($\Delta\omega = \omega_{LO} - \omega_{TO}$) of $\alpha$-quartz at selected bulk pressures. The group theoretical mode assignments are indicated. The E ($A_2$) phonon modes are polar with macroscopic electric field perpendicular(parallel) to the *c*-axis, respectively.

| | P=.24 GPa $\omega_{TO}$ | P=.24 GPa $\omega_{LO}$ | P=.24 GPa $\Delta\omega$ | P=7 GPa $\omega_{TO}$ | P=7 GPa $\omega_{LO}$ | P=7 GPa $\Delta\omega$ | P=19 GPa $\omega_{TO}$ | P=19 GPa $\omega_{LO}$ | P=19 GPa $\Delta\omega$ | P=24 GPa $\omega_{TO}$ | P=24 GPa $\omega_{LO}$ | P=24 GPa $\Delta\omega$ | P=30 GPa $\omega_{TO}$ | P=30 GPa $\omega_{LO}$ | P=30 GPa $\Delta\omega$ |
|---|---|---|---|---|---|---|---|---|---|---|---|---|---|---|---|
| E | | | | | | | | | | | | | | | |
| | 133 | 134 | 1 | 152 | 152 | 0 | 172 | 177 | 5 | 190 | 190 | 0 | 207 | 207 | 0 |
| | 263 | 265 | 2 | 294 | 298 | 4 | 341 | 349 | 8 | 352 | 366 | 14 | 347 | 380 | 33 |
| | 379 | 390 | 11 | 377 | 396 | 19 | 367 | 395 | 28 | 368 | 393 | 25 | 385 | 397 | 12 |
| | 445 | 500 | 55 | 470 | 520 | 50 | 497 | 544 | 47 | 505 | 552 | 47 | 514 | 562 | 48 |
| | 694 | 698 | 4 | 735 | 742 | 7 | 782 | 790 | 8 | 796 | 805 | 9 | 811 | 821 | 10 |
| | 795 | 807 | 12 | 834 | 845 | 11 | 876 | 883 | 7 | 888 | 893 | 5 | 899 | 902 | 3 |
| | 1050 | 1214 | 164 | 1049 | 1217 | 168 | 1048 | 1229 | 181 | 1048 | 1233 | 185 | 1047 | 1239 | 192 |
| | 1134 | 1129 | | 1124 | 1116 | | 1125 | 1112 | | 1125 | 1112 | | 1126 | 1125 | |
| $A_2$ | | | | | | | | | | | | | | | |
| | 343 | 368 | 25 | 334 | 363 | 29 | 334 | 373 | 39 | 336 | 379 | 43 | 337 | 387 | 50 |
| | 495 | 542 | 47 | 523 | 552 | 29 | 539 | 550 | 11 | 541 | 547 | 6 | 541 | 543 | 2 |
| | 765 | 787 | 22 | 770 | 820 | 50 | 774 | 857 | 83 | 776 | 868 | 92 | 780 | 882 | 102 |
| | 1063 | 1223 | 160 | 1080 | 1235 | 155 | 1103 | 1253 | 150 | 1110 | 1257 | 147 | 1116 | 1262 | 146 |

**Table V**: The computed long wavelength non-polar phonon frequencies ($cm^{-1}$) of $\alpha$-quartz at selected bulk pressures.

| | P=.24 GPa $\omega$ | P=7 GPa $\omega$ | P=19 GPa $\omega$ | P=24 GPa $\omega$ | P=30 GPa $\omega$ |
|---|---|---|---|---|---|
| $A_1$ | | | | | |
| | 239 | 291 | 303 | 306 | 310 |
| | 340 | 346 | 378 | 392 | 411 |
| | 465 | 516 | 574 | 591 | 608 |
| | 1068 | 1077 | 1098 | 1105 | 1118 |

**Table VI:** Zero pressure elastic constants (GPa) of?$\alpha$-Quartz (* $C_{12}=C_{11}-2C_{66}$).

| Elastic Constants | LDA calculations This work | LDA calculations Ref. [18]. | Experimental Ref. [5] | Experimental Ref. [3 ] |
|---|---|---|---|---|
| $C_{11}$ | 78.1 | 90 | 86.8 | 85.9 |
| $C_{12}^{*}$ | 16.0 | 12 | 7.04 | 7.16 |
| $C_{13}$ | 13.9 | 21 | 11.91 | 10.94 |
| $C_{14}$ | -15.7 | -12 | -18.04 | -17.66 |
| $C_{33}$ | 110.8 | 97 | 105.75 | 89.59 |
| $C_{44}$ | 55.2 | 61 | 58.2 | 57.66 |

**Table VII** : The computed unit cell constants and Si-O bond lengths in the monoclinic *C2* structure. The fractional coordinates x, y and z of the Si and O atoms in the asymmetric unit cell are listed. The Si1 silicon atoms are tetrahedrally coordinated, while the two equivalent Si2 atoms are in octahedral coordination; bond-length numbers in parenthesis represent the multiplicity. The zero pressure *C2* structure is metastable and has a higher free energy than the α-quartz structure. P* represents the structure under uniaxial pressure conditions with stress tensor $\sigma_{xx}$=$\sigma_{yy}$=0 GPa, $\sigma_{xy}$= $\sigma_{yz}$= $\sigma_{zx}$=0 and $\sigma_{zz}$= -16 GPa.

| | P=0 GPa | P=8.8 GPa | P=38 GPa | P* GPa |
|---|---|---|---|---|
| a (Å) | 8.5527 | 8.46909 | 8.2318 | 8.6327 |
| b (Å) | 3.4566 | 3.4200 | 3.3285 | 3.5142 |
| c (Å) | 5.375 | 5.29399 | 5.1040 | 5.1578 |
| β | 109.26147 | 109.85238 | 110.05966 | 111.05422 |
| Si1 (x,y,z) | 0., 0.5414, 0 | 0., 0.5423, 0. | 0., 0.5460, 0 | 0.,0.5548, 0. |
| Si2 (x,y,z) | 0.1797, 0.3086 , 0.6240 | 0.1774, .30818, .62029 | 0.1754, 0.3065, 0.6192 | 0.1762, 0.3068, 0.6157 |
| O (x,y,z) | 0.9897, 0.3212, 0.2734 | 0.9928, 0.3175, 0.2758 | 0.9975, 0.3081, 0.2768 | 0.9931, 0.3172, 0.2756 |
| O (x,y,z) | 0.8106,0.8084, 0.4279 | 0.8123, 0.8082, 0.4277 | 0.8160, 0.8070, 0.4268 | 0.8099, 0.8063, 0.4246 |
| O (x,y,z) | 0.1647, 0.7943, 0.0924 | 0.1656, 0.7982, 0.0893 | 0.1671,0.8086, 0.08114 | 0.1654, 0.7955, 0.0841 |
| Si(1)-O (Å) | 1.6831(2), 1.5915(2) | 1.6694(2), 1.5832 (2) | 1.6258(2), 1.5611(2) | 1.6682(2), 1.5805(2) |
| Si(2)-O (Å) | 1.6614, 1.7116, 1.7562, 1.7578 , 1.7730 ,2.0425 | 1.6564, 1.7087, 1.7352, 1.7354, 1.7712, 1.9584 | 1.6312, 1.6809, 1.6836, 1.6875, 1.7559,1.8555 | 1.6573, 1.7434, 1.7662, 1.7768, 1.7802, 1.8936 |

**Table VIII:** Comparisons of the computed *d*-spacings in the computed *C2* structure with the reported *d*-spacings in the observed post-quartz structures of Haines *et al.*[4] and Kingma *et al.*[1(b)].

| *C2* structure (P=38 GPa) *d*-spacing (Å) | Post-Quartz structure of Haines *et al.*[4] *d*-spacing (Å) | Post-Quartz structure of Kingma *et al.*[1(b)] *d*-spacing (Å) |
|---|---|---|
| 4.794 | | |
| 3.867 | | 3.776 |
| 3.691 | 3.603 | 3.613 |
| 3.057 | 3.161, 2.976 | 3.027 |
| 2.751 | 2.745 | 2.668 |
| 2.605 | | |
| 2.448 | 2.48 | |
| 2.397 | 2.225 | |
| 2.091 | 2.175 | 2.129 |
| 2.024 | 1.968 | |
| 1.846 | 1.926 | |
| 1.782 | | |
| 1.716 | 1.725 | |
| 1.665 | | |
| 1.572 | 1.546 | |
| 1.529 | 1.517 | |
| 1.513 | 1.492 | |
| 1.344 | 1.413 | |
| 1.302 | 1.305 | |
| 1.275 | 1.278 | |
| 1.275 | 1.258 | |
| 1.216 | 1.236 | |
| 1.188 | 1.191 | |
| 1.171 | 1.169 | |

Fig. 1 : (Color online). Polyhedral representation of the computed α-quartz and the high pressure *C2* structure obtained using the software xtaldraw[52]. The crystallographically distinct silicon atoms are shown with different shades.

a **-Quartz**

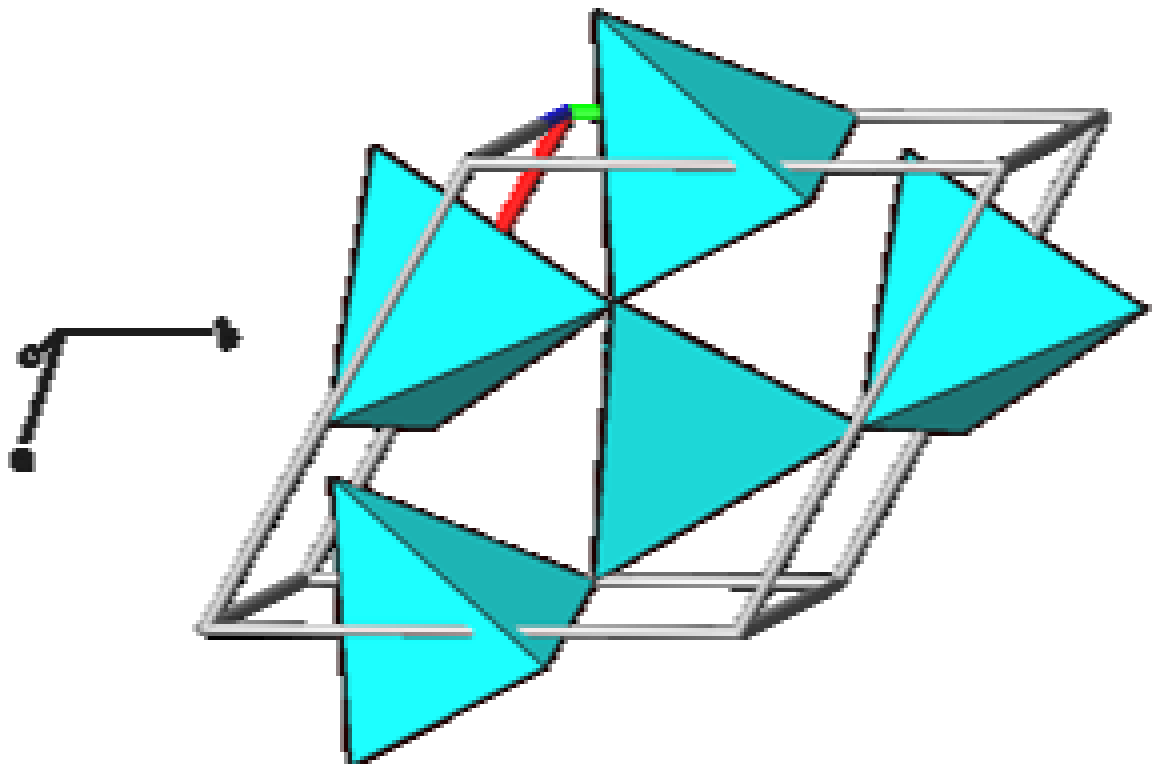

***C2*** **Structure**

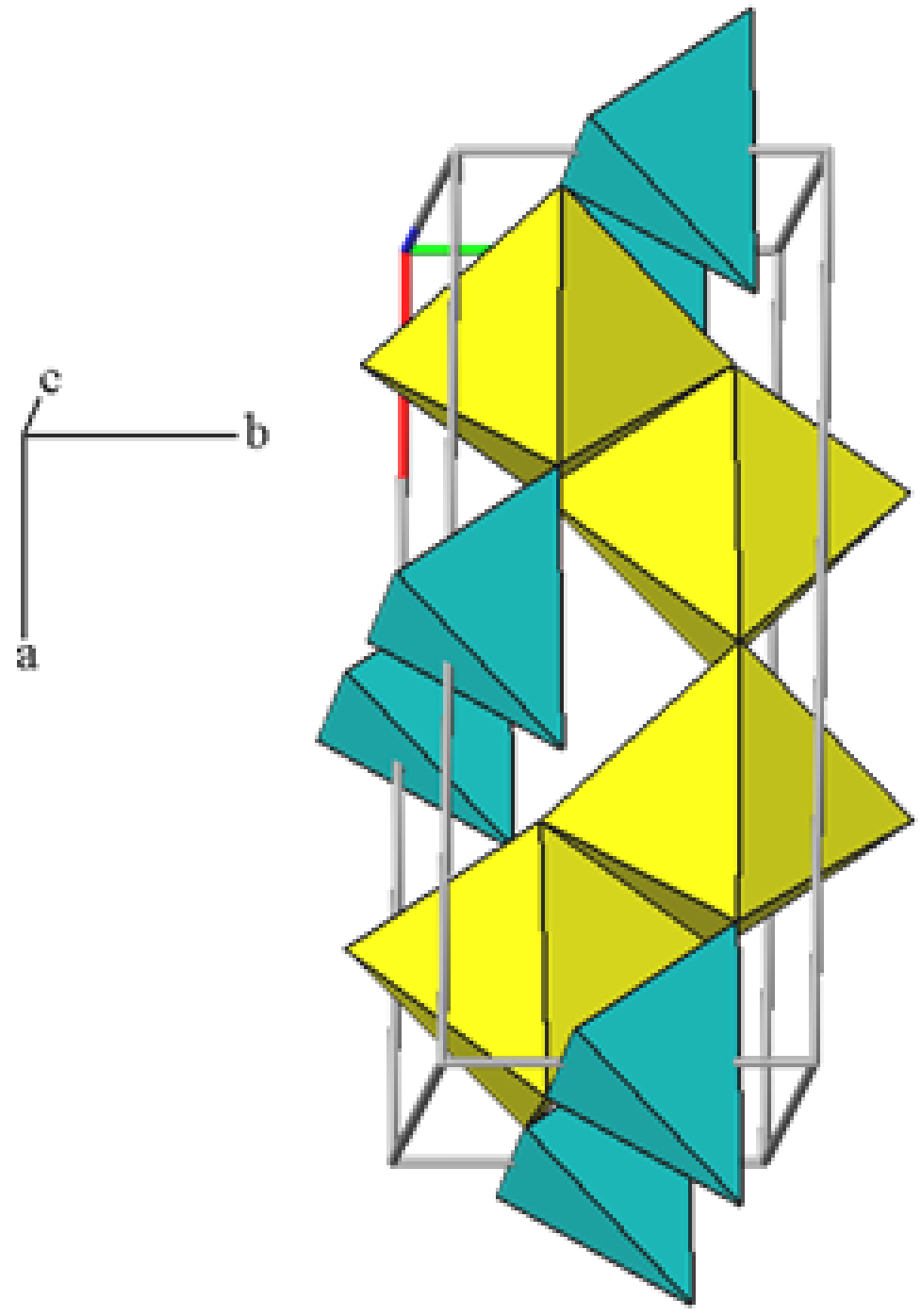

**Fig. 2**: (Color Online): Comparison of the calculated (full line) equation of state of α-quartz with reported LDA and GGA calculations [49] and experimental data[28(a-d)](symbols).

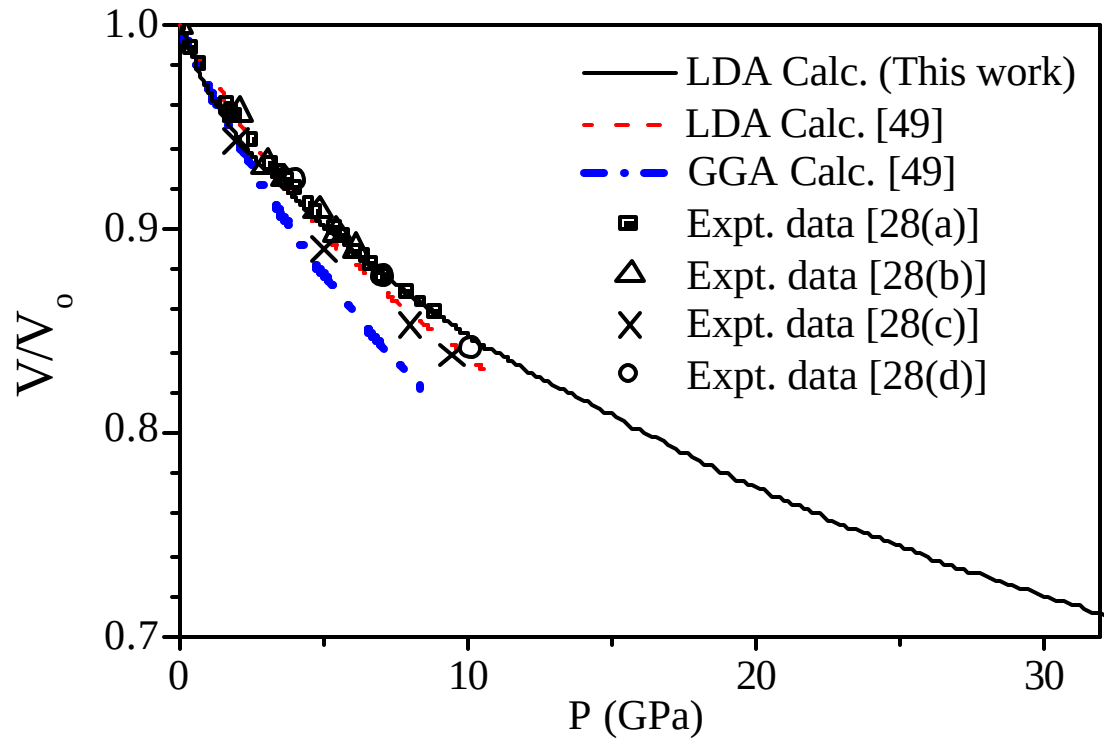

**Fig. 3**: Comparison of the computed (open symbols and lines) and experimental[29] (filled symbols) long wavelength phonon frequencies as a function of hydrostatic pressure. The group theoretical phonon mode assignments are indicated.

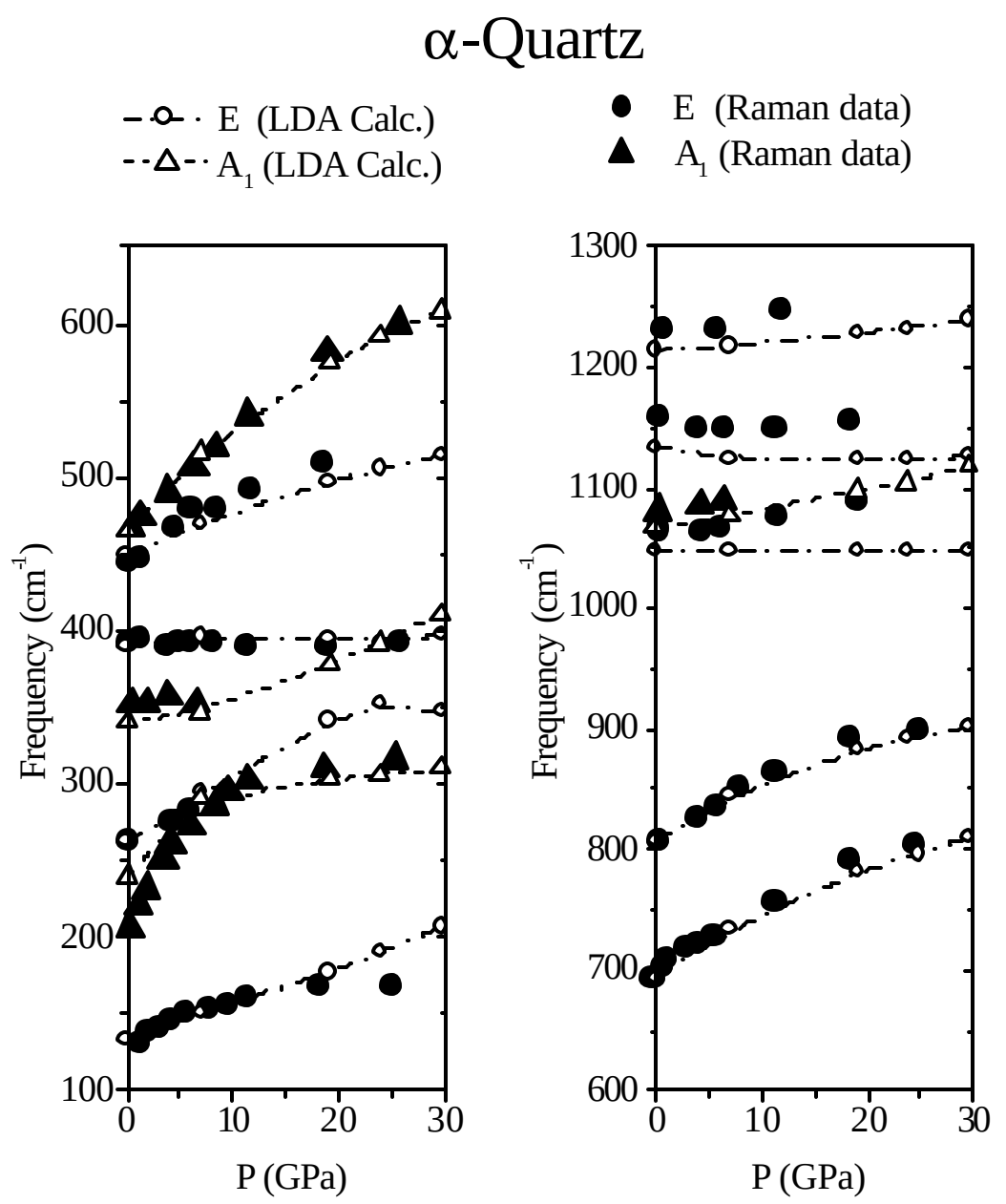

**Fig. 4.** Comparison of the computed elastic constants as a function of pressure with the reported Brillouin scattering single crystal experimental data of Gregoryanz *et al.* [3(a)] .

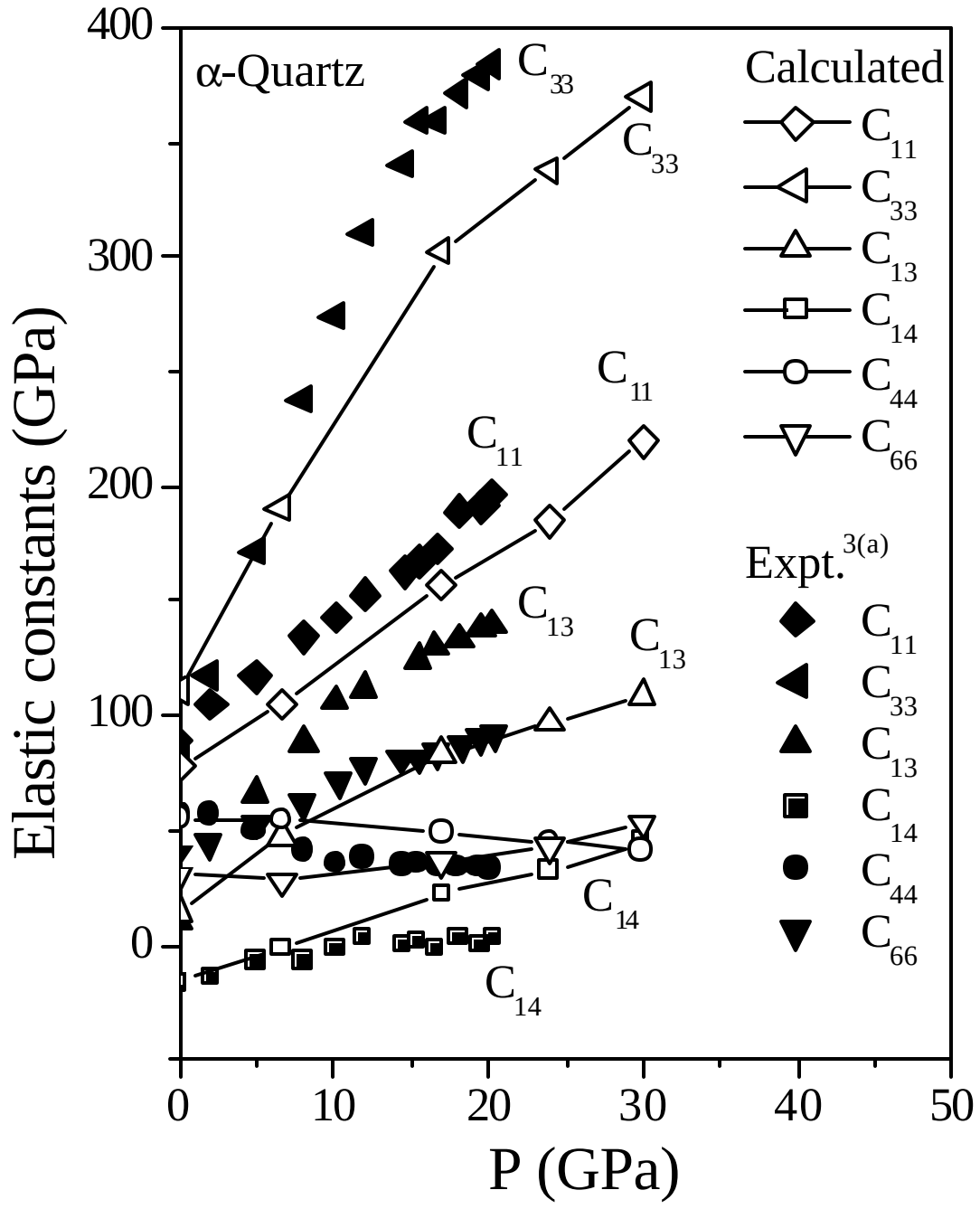

**Fig 5.** The variations of the computed $B_1$, $B_2$ and $B_3$ and the soft acoustic mode elastic constant $rv^2$ with pressure. $B_3$ and $rv^2$ are negative above P~32 GPa, which indicates the onset of mechanical instability in α-quartz at high pressure. In (d), we also display the variation of the calculated soft mode K-point phonon frequency with pressure.

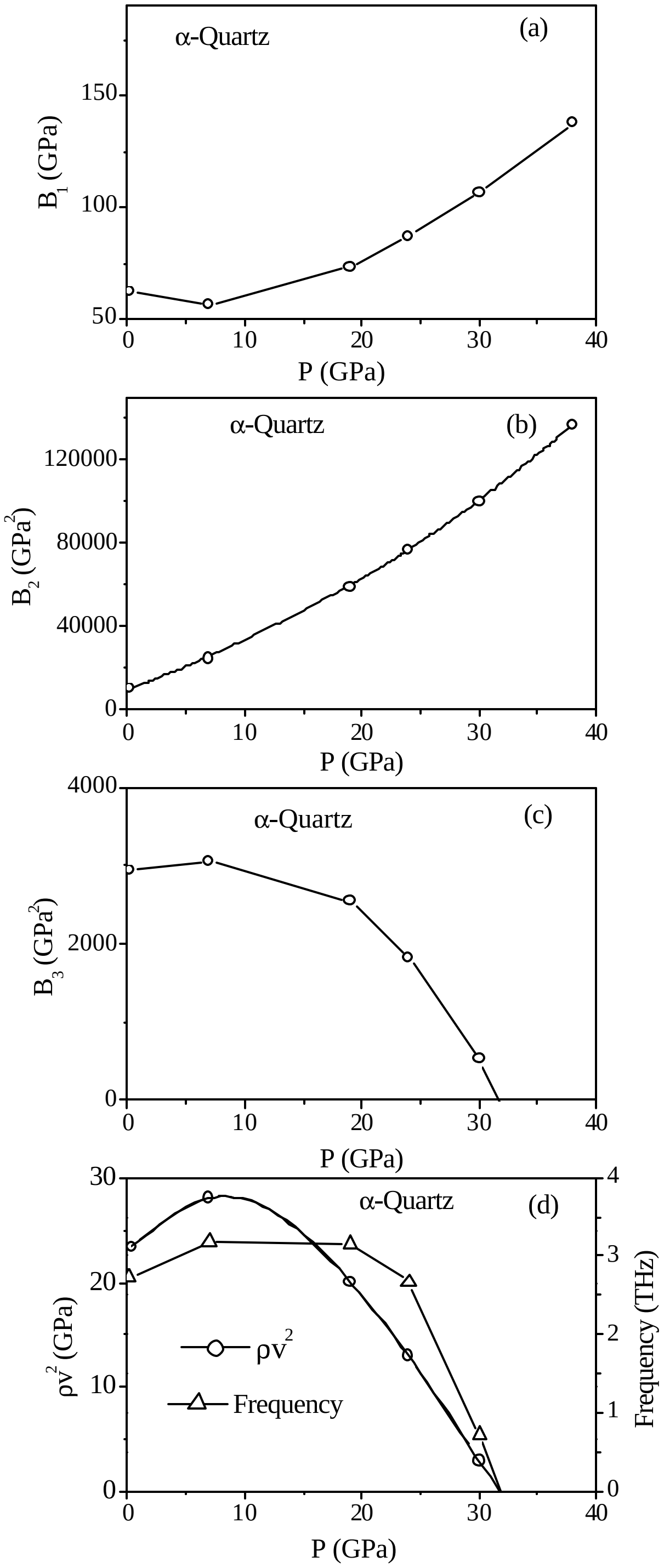

**Fig. 6.** The computed high pressure phonon dispersion relations of α-Quartz. While Γ is the zone center point, K, M and A are zone boundary points in the Brillouin zone[9]. The wave vector directions are as indicated on the top of the figure in (a). At high pressures, the lowest phonon mode at the K-point becomes unstable. While (a), (b), (c) and (d) give the phonon dispersion of α-Quartz for bulk hydrostatic pressures, (e) corresponds the non-hydrostatic situation with $\sigma_{xx}=\sigma_{yy}=-19$ GPa, $\sigma_{xy}=\sigma_{yz}=\sigma_{zx}=0$ and $\sigma_{zz}=-29.6$ GPa. (f) has the stress tensor $\sigma_{xx}=\sigma_{yy}=0$ GPa, $\sigma_{xy}=\sigma_{yz}=\sigma_{zx}=0$ and $\sigma_{zz}=-15.8$ GPa.

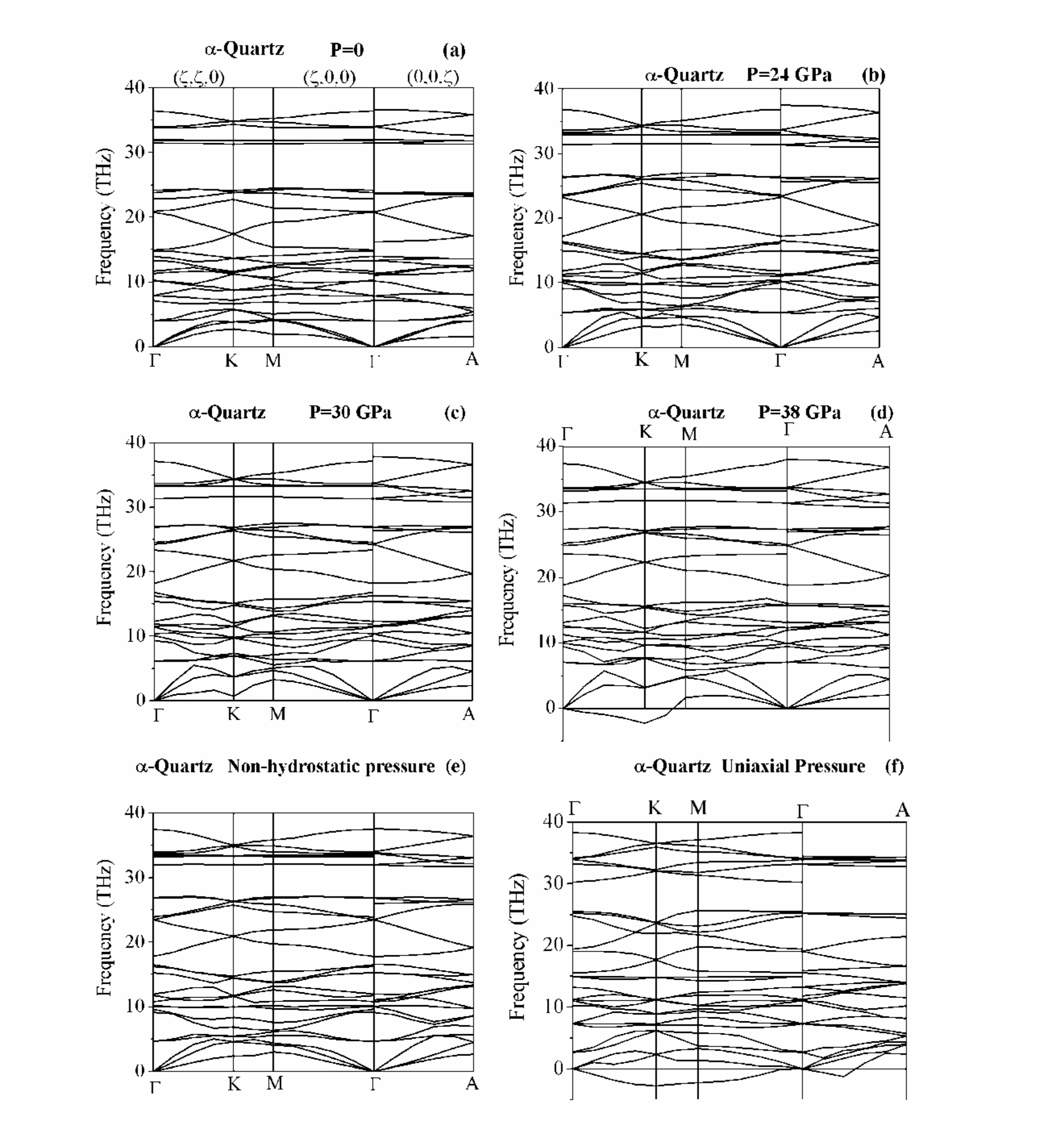

**Fig. 7.** The computed enthalpy of α-Quartz and the *C2* structure as a function of pressure.

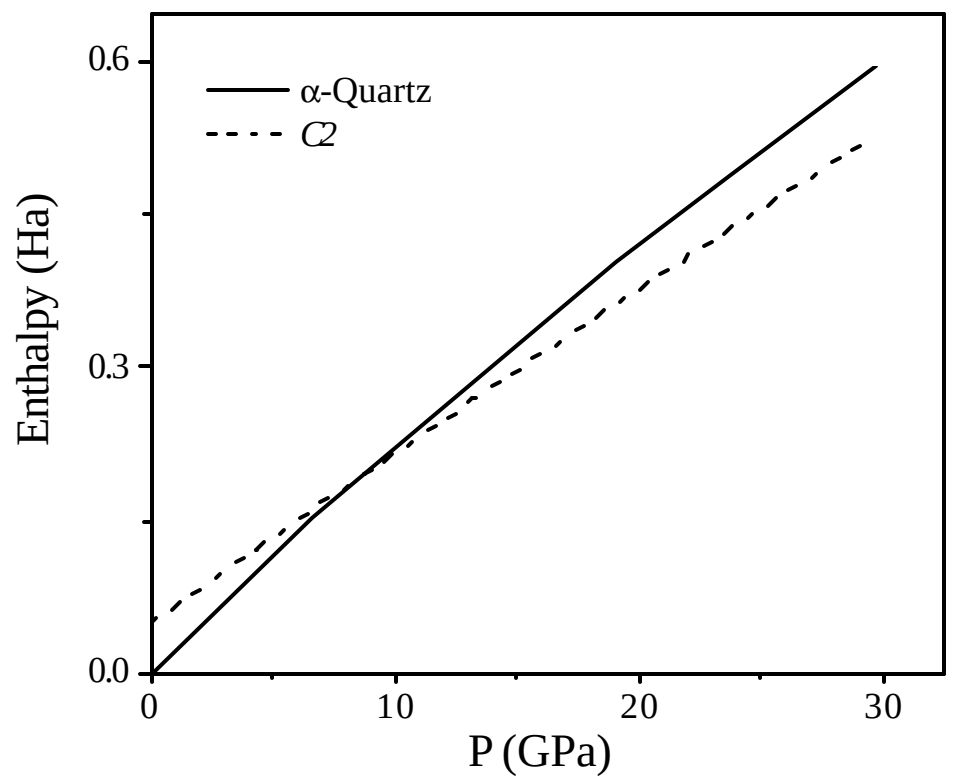

**Fig. 8.** Comparison of the observed x-ray diffraction intensities[4] in the reported (λ=0.41693 Å) post-quartz phase of Haines *et al*.[4] (a) with the computed x-ray intensities in the *C2* structure (b). Peaks labeled Q in (a) correspond to quartz peaks. The calculated *d*-spacings and energy dispersive spectra of the *C2* structure (c-d) are compared with the reported synchrotron energy-dispersive x-ray diffraction pattern (e) in the post-quartz phase crystalline phase of Kingma *et al*.[1(b)]. In (e), * represents new post-quartz phase peaks, Q' are shifted quartz-like diffraction peaks and Ne has been identified as diffraction lines of the pressure transmitting medium neon[1(b)].

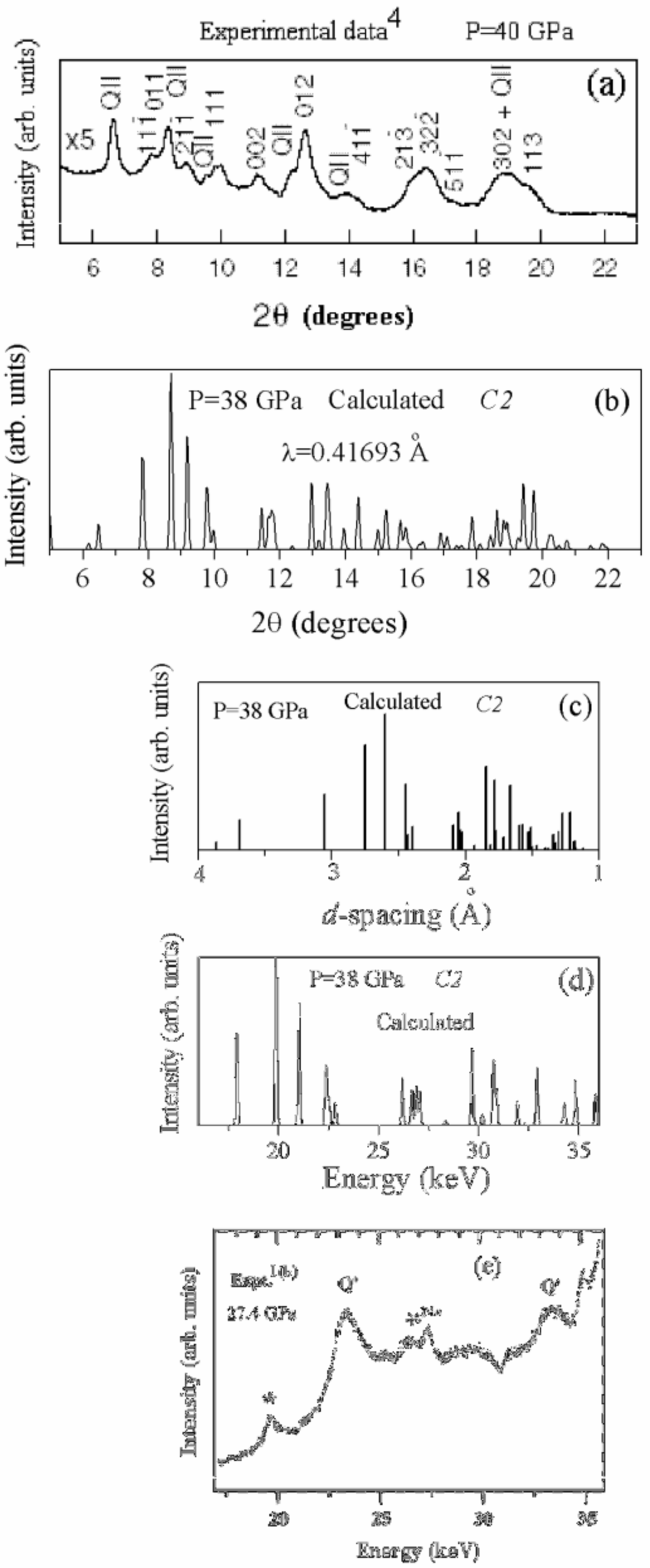